# Sodium Doped LaMnO$_3$ Thin Films: Influence of Substrate and Thickness on Physical Properties

*Lorenzo Malavasi\*[a], Maria Cristina Mozzati[b], Ivano Alessandri[c], Laura E. Depero[c], Carlo B. Azzoni[b] and Giorgio Flor[a]*

[a]Dipartimento di Chimica Fisica "M. Rolla", INSTM, IENI/CNR Unità di Pavia of Università di Pavia, V.le Taramelli 16, I-27100, Pavia, Italy.
*E-mail*: lorenzo.malavasi@unipv.it

[b]INFM, Unità di Pavia and Dipartimento di Fisica "A. Volta", Università di Pavia, Via Bassi 6, I-27100, Pavia, Italy.

[c]INSTM, Laboratorio di Strutturistica Chimica, Dipartimento di Ingegneria Meccanica, Università di Brescia, via Branze 38, I-25123, Brescia, Italy.

## Abstract

In this paper we report the results about the synthesis and characterization of optimally doped La$_{1-x}$Na$_x$MnO$_3$ thin films grown onto SrTiO$_3$ (100), NdGaO$_3$ (100) and NdGaO$_3$ (110) for thickness ranging from 11 to 82 nm. The effect of substrate nature and orientation, film thickness and annealing procedure was investigated in order to optimize their magnetoresistance (*MR*). We obtained very smooth films displaying *MR* values greater than 70%, near to room temperature.

# Introduction

The synthesis and characterization of manganite thin films have reached nowadays a relatively high level of expertise. To a great extent, these results have been possible thanks to the great technological improvement of oxides thin films research[1], mainly due to the discovery and interest grown in the past fifteen years in the high temperature superconductors field.

The renewed interest on the doped-manganites came out from the observation of a high magnetoresistive effect, designed as "colossal" (CMR), which triggered the attention of the scientific community both for the fundamental research and for the applicative possibilities, particularly for the spin-dependent transport devices[2-6]. Of course, thin films are the most suitable form for practical purposes.

The research efforts during last ten years have pointed out the role played in the manganites by the different oxidation states, 3+ and 4+, of the manganese ions: in $LaMnO_{3+\delta}$ and $La_{1-x}Ca_xMnO_3$ systems, for example, a $Mn^{4+}$ content around ~30% was found to be the optimal value in order to achieve the maximum CMR effect and also the highest Curie temperature ($T_C$) for the paramagnetic to ferromagnetic transition (P-F)[7-12]. In the magnetic ordered state the electron hopping process is favored by the spin alignment while for temperatures above $T_C$ the conduction mechanism is semiconducting-like. We stress that in the metallic-regime, the actual oxidation state of Mn ions is an intermediate one since the charge carriers become itinerant and a net oxidation can not be defined anymore for the Mn-ions.

One of the most exciting features peculiar to manganites thin films is the close correlation between the electronic transport and the strain effect induced by the substrate. In fact the magnetoresistance response (*MR*), the semiconducting- to metallic-like transition temperature ($T_{S-M}$), and P-F transition temperature ($T_C$) can be nicely tuned by controlling the film thickness and the substrate nature, *i.e.*, the two most important variables in affecting the film strain[13-16] and growth direction.

Most of the works published in the last years on manganites thin films focused on the La-Ca-Mn-O (LCMO) and La-Sr-Mn-O (LSMO) systems, particularly for historical reasons, since they were the first phases discovered and studied, especially as bulk materials. The LCMO system exhibits high *MR* values, around 60-70%, but at temperature usually lower than 250 K[14]; on the opposite, the LSMO system is metallic and ferromagnetic at higher temperatures (above 300 K) but displays lower *MR* values, often < 40% at 5 T[17]. So, the search for new phases able to display relatively high *MR* values at temperature closer to room-temperature (RT) is still an important task.

We recently focused on the sodium doped lanthanum manganites[18-21] in order to deeply investigate this system which looks suitable for achieving better electronic and magnetic properties. This is mainly due to the fact that the tolerance factor (*t*) ($t = (A\text{-}O)/\sqrt{2}(Mn-O)$ where A-O and Mn-O are the equilibrium bond lengths) is practically unchanged by the sodium replacement of lanthanum due to the likeness of ionic radii; moreover, compared to the Ca-doped manganites, it is possible to achieve an equal amount of hole doping with a lower cation substitution. This should reflect in a lower cation disorder induced by the doping. Finally, for an optimal doping, these materials present a rhombohedral structure which is more favorable for the transport properties with respect to the orthorhombic one usually displayed by the LCMO.

To our knowledge, in the current literature only two works can be found dealing with the preparation and characterization of sodium doped $LaMnO_3$ thin films[22,23]; moreover, none touches on the rf-sputtering deposition, which looks the most suitable technique for a scale-up and technological transfer of the laboratory results.

In this paper we report about the rf-magnetron sputtering deposition of $La_{0.85}Na_{0.15}MnO_3$ (LNMO) thin films onto three different substrates, $SrTiO_3$ (100), $NdGaO_3$ (100) and $NdGaO_3$ (110), with different films thickness. The properties of the deposited films were studied by means of X-ray diffraction (XRD), X-ray Reflectivity (XRR), Atomic Force Microscopy (AFM), resistance (*R*) measurements with and without applied magnetic fields. Electron Paramagnetic Resonance (EPR) and magnetization (*M*) measurements were performed only on the films grown on SrTiO$_3$ (STO) due to the paramagnetism of the NdGaO$_3$ (NGO) substrate ($J$(Nd$^{3+}$)=9/2).

**Experimental**

La$_{0.85}$Na$_{0.15}$MnO$_3$ thin films were grown starting from powdered target material synthesized by solid state reaction; details about the target preparation and characterization are reported in Ref 18. Thin films were deposited on SrTiO$_3$ (100), NdGaO$_3$ (100) and NdGaO$_3$ (110) single crystals (Mateck®). SrTiO$_3$ single crystals are cubic (space group P*m-3m*) with a lattice parameter *a* = 3.90 Å while NdGaO$_3$ is orthorhombic (space group P*bnm*) with *a* = 5.43 Å, *b* = 5.50 Å and *c* = 7.71 Å. The depositions were performed by an off-axis rf-magnetron sputtering system (RIAL Vacuum®); the gas in the chamber was an argon/oxygen mixture in the ratio 12:1. This gas composition was chosen since the formation of cation stoichiometric films with the correct A (La, Na)/Mn ratio, with respect to the target material, can be well accomplished in an oxygen-poor gas environment, as observed in an our previous investigation[19]. The total pressure in the sputtering chamber was 4•10$^{-1}$ Pa and the rf power was kept at 145 W. The substrate temperature, measured by a K-type thermocouple located under the substrate, was set at 973 K. Film thickness was monitored by means of an internal quartz microbalance and defined by X-ray reflectivity measurements (XRR). To ensure a complete oxygen stoichiometry, as well as to improve the sample crystallization, an oxygen *ex situ* annealing was carried out at 1173 K for 2 hours in a quartz home made apparatus in which a certified gas mixture was flown; at the end of the thermal treatments the samples were quenched to RT.

Correct cation stoichiometry was also indirectly confirmed by X-ray absorption spectroscopy measurements[24] performed at the Mn-K edge which showed that the energy position of the edge was the same for all the thin films and the bulk material. Moreover, since for the bulk material the oxygen content was accurately measured[18] to be 3, we are able to state that also the oxygen stoichiometry of the samples should be very close to this value otherwise sensible variation of the Mn-K edge should have been detected. Finally, the same energy position means also analogous $Mn^{3+}/Mn^{4+}$ ratios for all the samples which is around 30% as previously determined for the bulk phase[18]. This allows us to correlate all the results of the physical characterization only to the role of the film orientation, strain and thickness.

XRR and $\theta$-$2\theta$ XRD data were collected by using a Bruker "D8 Advance" diffractometer equipped with a Göbel mirror. The Cu $K_\alpha$ line of a conventional X-ray source powered at 40 kV and 40 mA was used. All the reflectivity spectra were analyzed with the REFSIM software package[25].

NC-AFM images were obtained with an Autoprobe CP microscope (Park Instruments-VEECO), operating in non-contact mode, by means of silicon tip onto V-shaped cantilevers (resonance frequency: 120 KHz; force constant: 3.2 N/m). Standard $2^{nd}$ order flatten processing of the images has been performed in order to correct the scanner non-linearity.

Resistance measurements were carried out in a Quantum Design Physical Properties Measurement System (PPMS) in a four points configuration. The electrodes were glued on the film with an epoxidic silver paint (Epoteck®).

A SQUID magnetometer (Quantum Design) was employed for static magnetization measurements. *M* has been measured as a function of temperature and magnetic field (*H*) in the temperature range 2 – 350 K and for fields ranging between 0 and ±10000 Oe. The sample holder (plastic straw) and all the diamagnetic contributions have been taken into account.

EPR measurements were performed at ~ 9.5 GHz with a Bruker spectrometer equipped with a continuous nitrogen flow apparatus in the temperature range between 140 and 400 K.

We stress that the last two experimental techniques were applied only for the study of thin films deposited on STO due to the strong paramagnetic signal coming from the $Nd^{3+}$ present in the NGO substrates.

**Results and Discussion**

*1. Structural characterization*

La$_{0.85}$Na$_{0.15}$MnO$_3$ starting powders present a rhombohedral structure belonging to the $R\bar{3}c$ space group; cell parameters, obtained from the refinement of the hexagonal unit cell, are the following: $a = b = 5.494$ Å, $c = 13.302$ Å, V = 347.76 Å$^3$ [18].

Thin films were grown on STO (100) and NGO (110) and (100) for increasing sputtering times from 4 to 30 minutes. The films thickness was carefully measured by XRR. From the results of the reflectivity characterization, reported in Table 1, a growth rate around 2.6 nm/min was estimated.

The $\theta$-$2\theta$ XRD patterns acquired on the prepared thin films allowed to check the presence of only those reflections related to the nature of the substrate, *i.e.* the ($h$00) for the STO (100), ($h$00) for NGO (100) thin films and the ($hk$0) for the NGO (110) thin films. Moreover, in order to separate the contribution of the film from that of the substrate, further XRD spectra have been carried out with an unspecular setting, i.e. with the detector and the incident beam not exactly at the same angle. We stress that in the following discussion of the XRD data the pseudocubic notation is used for the thin films.

In Figure 1a is reported, as a selected example, the XRD $\theta$-$2\theta$ scan for the 13 nm LNMO thin film on STO, while in Figure 1b is reported a scan for the LNMO grown on NGO (110) of thickness 82 nm. Finally, Figure 1c reports the $\theta$-$2\theta$ scan for a film of the same thickness grown onto NGO (100).

The in-plane lattice constants mismatch between the STO and LNMO target material, defined as ($a_{substrate}$-$a_{target}$/$a_{substrate}$)•100, is around +0.50% (with reference to the pseudocubic cell), so a tensile strain, that is a decreasing in the growth direction and expanding in the plane, is expected for this films series. On the opposite, for the films grown onto NGO (110) the mismatch is about -0.54% thus giving origin to a compressive strain of about the same order of that present for the STO substrate but of opposite direction; in this case a small compression of the in-plane parameters and an elongation of the out of plane one is expected. Finally, for the films grown on the NGO (100), even though we have not a direct measurement of the in-plane orientation, based on the position of thin films reflections, the oriented growth should place the *b-c* lattice parameters of the orthorhombic *Pbnm* structure in the plane of the substrate while the out-of-plane lattice constant will be one of the two short axes of the orthorhombic cell (*a*, in this case).

The lattice constants determined by the XRD characterization confirmed the opposite nature of the strain for the films deposited on STO (100) and NGO (110), also suggesting a progressive decreasing of the stress by increasing the thin film thickness as usually observed[14,17,26]. The out-of-plane lattice parameter for the films grown onto STO (100) and NGO (110), the *c*-axes of the pseudo-cubic unit cell, is reported in Table 1. As can be appreciated for the STO-series, the *c*-parameter increases passing from 13 to 25 nm and then it remains practically constant, while, on the opposite, for the NGO (110)-series the parameter progressively reduces with thickness increasing. On the other hand, the films grown onto NGO (100) substrate are forced to experience a highly anisotropic situation due to the appreciable difference existing between the orthorhombic parameters exhibited by the (*h*00) planes of the substrate.

An AFM characterization was performed on all the prepared films, before and after the thermal treatments. For all of them a general and common morphology was observed. However, in the following we will report only the results for a representative sample with the intention that the results found for this sample are of general meaning. Figure 2 reports, as a selected example, the AFM images taken on an as-prepared LNMO thin film of 33 nm grown onto NGO (110). Figure 2a and 2b reports the 2D-images collected for a surface of 2.8x2.8μm and 0.5x0.5μm, respectively.

The correspondent 3D-images are presented in Figures 2c and 2d. The film is made of small grains with average dimension lower than 20 nm. The average roughness, defined as:

$$R_{ave} = \sum_{n=1}^{N} \frac{|z_n - \bar{z}|}{N}. \tag{1}$$

is around 0.28 nm for the 0.5x0.5μm image (2).

The surface plots for the same film but annealed are reported in Figures 3 (a-d) for areas of 5x5μm (3a), 1.7x1.7μm (3b) and 0.7x0.7μm (3d). The 3D-image for the 1.7x1.7μm area is reported in Figures 3c. From Figure 3a it is possible to appreciate that the thin film is constituted of grains of mean size around 150 nm with a rounded and slightly elongated shape. If a smaller area is considered (Figure 3b) it is possible to better appreciate the grains morphology; moreover, it is clearly notable the presence, beside the bigger grains, of smaller particles, with sharing boundaries, with dimensions lower than 100 nm. Finally, looking in more detail at the single grain features (Figure 3d), even though the image quality is lower with respect to the previous ones, it is possible to notice that the grains, which in the precedent figures looked as single particles, are actually made of very small grains sintered together of mean size around 50-60 nm. So, during the annealing step the size and shape of the grains drastically changed. Grain growth is accompanied by a rounding effect due to the surface energy minimization. For the range of thickness explored in the present paper we did not observe a marked dependence of grain size with film thickness which instead was clearly observed when spanning greater thickness[27]. Concerning the roughness, the annealing treatment causes an increase of it giving a value of $R_{ave}$ around 1.3 nm for the 0.7x0.7μm surface. These data also suggest that the growth of film, both during deposition and during annealing, is in the *a-b* plane and that it starts from small nuclei with sizes comparable to the film thickness which then grow in the plane by coalescing one with the other. The thermal treatments allow the formation of big island of hundred nanometers.

*2. Magnetoresistance*

Figure 4 reports the resistance ($R$) versus temperature ($T$) plots for the annealed films of LNMO on STO (100) of thickness 13 nm (a), 25 nm (b) and 70 nm (c), respectively. The resistance measurements were carried out at null applied magnetic field (solid lines) and with an applied magnetic field of 7 T (dash-dot lines). The correspondent MR, defined as

$$MR(\%) = \frac{R(H) - R(0)}{R(0)} \bullet 100 \qquad (2)$$

is also presented in the graphs (empty circles).

Looking at the $R$ vs. $T$ curves without applied magnetic field it is possible to note the presence of a transition from a semiconducting-like (S), activated, regime towards a metallic (M) transport for all the samples. For films of 13 and 25 nm, one S-M transition occurs, while for the thicker film two distinct transitions are present. The $T_{S-M}$ values, taken at the maximum of the $R$ vs. $T$ curves, are reported in Table 1 and pass from 248 K for the 13 nm film to 260 K for the 25 nm one. The two $T_{S-M}$ for the 70 nm film are found at 260 and 182 K, respectively. In addition, it is also possible to appreciate that, passing from the 13 to the 25 nm thin film, not only the $T_{S-M}$ increases but also the width of the transition reduces.

The $R$ vs. $T$ curves at 0 T were further analyzed for what concern the nature of the charge carrier transport, particularly in the activated regime, modeled considering a small polaron conductivity according to:

$$R = R_0 T \exp(E_a / k_B T), \qquad (3)$$

which proved to nicely describe the resistance behavior for the LCMO system[9,28,29]. The activation energies ($E_a$), as obtained by the fitting procedure, are reported in Table 1 for all the considered films.

As can be seen, for the films grown onto STO (100), the $E_a$ values are nearly equal for the 13 and 25 nm films (around 95 meV) and then $E_a$ increases to 129 meV for the thicker film (70 nm). In this case we considered the activated part concerning with the first transition placed at 260 K.

The application of a magnetic field of 7 T induces in all the LNMO thin films grown onto the STO (100) substrate a reduction of the resistance and a shift at higher temperatures of the $T_{S-M}$ values. The resistance change is well evidenced by the *MR* curves according to expression (2). The temperature values of the *MR* maxima and their absolute values are also listed in Table 1 for all the samples. The *MR* is around 80% for the 13 and 25 nm thin films with its maxima placed at 240 and 252 K, respectively. For the 70 nm film the two *MR* values are 45% at 259 K and 37% at 178 K.

Figures 5a, 5b and 5c report the same kind of measurements, *R vs. T* at $H = 0$ (solid lines), $H = 1$ (dashed line) and $H = 7$ T (dash-dot lines), carried out on the series of LNMO thin films deposited onto NGO (110). In this case the thin film thickness are: 11 nm (5a), 33 nm (5b) and 82 nm (5c).

The first film does not display any transition and the transport is always activated in nature. Moreover, the resistance value for this sample is extremely high ($>10^5$ Ω) thus appreciably reducing the investigable temperature range. On the opposite, for the two other samples a S-M transition is encountered at 285 and 256 K, respectively, in the $H = 0$ curves (solid lines). Moreover, the width of the transition is nearly the same for both samples.

Also in this case we estimated the $E_a$ values for the hopping of charge carriers between $Mn^{3+}$ and $Mn^{4+}$ ions. The highest $E_a$ pertains to the thinner sample (155 meV) and the lowest one (71.6 meV) to the 30 nm film.

The application of a magnetic field causes the drop of resistivity in those samples displaying a S-M transition. For the 30 nm one (Figure 5b) the *MR* peak at 1 T (dashed line) is located at 262.5 K with a *MR* absolute value of about 26% while the application of a 7 T magnetic field moves the *MR* peak to 271 K with a maximum resistance reduction around 70%. The sample of thickness 82 nm (Figure 5c) presents an *MR* peak located at lower temperatures, in accordance with the lowest $T_{S-M}$ values: at 1 T the *MR* peak is found at 233 with a *MR* value of 37% and shifts to 240 K with a maximum *MR* value greater than 91% when the field is raised to 7 T.

Let's consider, finally, the thin films series grown onto NGO (100). In this case we could not find any evidence of transition in the *R vs. T* plots for the three studied samples. For all the considered thickness, *i.e.* 11, 33 and 82 nm, the transport is always semiconducting-like and the samples show very high resistance values. The *R vs. T* curves for the three thin films are presented in Figure 5d (the temperature range of measurement was reduced as a consequence of the high resistivity of the samples). The application of a magnetic field of 7 T does not decrease the resistance of the samples apart for a small reduction observed under 200 K for the thicker sample. According to the *R vs. T* behavior these samples show the highest activation energies which progressively diminish by increasing the film thickness. The values of $E_a$ for the three samples are reported in Table 1.

As a general comment on the activation energies for all the three series, from Table 1, it is clear there is not a direct correlation between the thickness and the $E_a$ values but rather with the strain effect. Moreover, the lowest energy required for the polaron hopping is achieved for the film with the highest $T_{S-M}$ value.

*3. Magnetic Properties*

The charge transport in manganites is closely coupled to their magnetic state. Zero field cooled (ZFC) and field cooled (FC) *M vs. T* curves have been obtained by applying a 1000 Oe magnetic field to the films grown on STO. Figure 6 shows these curves for the samples with *H* perpendicular to the out-of-plane axis (*c*). All the samples show a wide transition with a shape observed for this kind of materials[30,31]: *M* first increases rapidly and then slowly with decreasing the temperature. The $T_C$ values, taken at the beginning of the *M* rapid increase, monotonically decrease by increasing the films thickness, being about 275, 255 and 215 K for the 13, 25 and 70 nm film, respectively. The thinnest sample also shows the highest $M$ values for $T < T_C$. The small difference between the ZFC and FC curves at very low temperatures suggests the absence of magnetic clusters.

We remember that the $M$ vs. $T$ curve of a bulk polycrystalline sample, with composition and oxygen content analogous to the films here considered (and so with a Mn(IV) amount around 30%), shows a sharp $M$ increase, correspondent to a P-F homogeneous magnetic transition at about 300 K[18]. On the contrary, the gradual $M$ increase when $T$ is decreased, observed here for the films, looks rather more similar to the behavior observed for nanosized polycrystalline samples, prepared *via* propellant synthesis and with estimated grain dimension around 35 nm, and ascribed to grain size distribution and unhomogeneity (as also deduced from TEM measurements)[20]. In these samples a superparamagnetic behavior was observed, since no saturation occurs up to the lowest investigated temperature (2 K) and the highest magnetic field (7 T), also according to EPR data[20]. On the contrary, for the films the magnetization saturation is reached at 2 K by applying very low magnetic fields, so suggesting a P-F transition.

The onset of a ferromagnetic order below $T_C$ is indeed proved by the hysteresis cycles. In Figure 7 the $M$ vs. $H$ curves at 2K of the films are reported, showing that $M$ rapidly increases for $H < 500$ Oe and then reaches its saturation value, with almost rectangular hysteresis loops, as better evidenced in the inset. So, the ferromagnetic order, not observed in the polycrystalline nanosized materials, results, for the films, from the grains orientation itself. At 2 K all the samples show similar values of coercive field and magnetization saturation. The net magnetic moment per Mn ion, as deduced by the saturation values, is about 2.5 $\mu_B$, so lower than the expected one (about 3.7 $\mu_B$[32], by considering that the $Mn^{3+}/Mn^{4+}$ ratio is the same in all the analyzed samples and around 2.3).

This is consistent with small sized particles in agreement with the morphological characterization. Indeed a broadening of the P-F transition and a decrease of $T_C$ and $M$ values, with respect to the bulk samples, is expected[33] from the reduced particle size and the higher grain boundary contribution. Only for a 150 nm thin film (not considered in the paper) a net magnetic moment close to the expected value was achieved (about 3.4 $\mu_B$).

The magnetic properties of these manganites are extremely sensitive to structural changes or disorder and also the lattice strain induced by the underlying thin film substrate can have an important influence[34,35]. Greater Mn-O-Mn bond angles and smaller Mn-O bond lengths enhance $M$ and $T_C$ values because the exchange interaction between the Mn ions becomes stronger[33,34]. For our films the larger in-plane lattice parameters of STO substrate elongate the $a$ and $b$ manganite lattice parameters so increasing the distance between Mn ions with respect to the bulk material, with a general reduction of $M$ and $T_C$ values. Besides, the STO cubic structure makes the Mn-O-Mn angle closer to 180°, so enhancing the ferromagnetic exchange interactions. Of course the thinnest films are expected to be more uniformly strained[36] and by increasing the film thickness a non homogeneous decreasing of the strain (above a critical thickness) has to be taken into account[35]. This mainly explains the highest $T_C$ and $M$ values pertaining to the thinnest film.

The good magnetic homogeneity suggested by the almost rectangular hysteresis loops and by the very similar ZFC and FC $M$ vs. $T$ curves is also confirmed by EPR measurements. Indeed the EPR spectra consist of a unique narrow signal component at any investigated temperature, as shown in Figure 8 for the 13 nm thin film; this reflects the magnetic and chemical homogeneity of the investigated samples, as already observed for analogous nanosized polycrystalline materials prepared via sol-gel route[20]. So, the wide P-F transition of our thin films is not ascribable to microscopic magnetic unhomogeneity across the films but, rather, to the small particle size, which may determine a distribution of the exchange coupling strength arising from the weaker magnetic interaction near the grain boundary, compared to the intra-grain contribution[33]. This effect is also consistent with the average magnetic moment of Mn ions lower than the 3.7 $\mu_B$, as instead expected for bulk samples.

The temperature dependence of the EPR resonant field when applying the magnetic field both parallel and perpendicular to the sample surface evidences the theoretical foreseen behavior within an approach based on the Bloch equations including the demagnetization term and no magnetic anisotropy[37]. This approach cannot well describe at all the trend of the EPR resonant fields observed for analogous thin films but grown an amorphous substrate (not shown here). In Figure 9 the temperature dependence of the EPR resonant field is reported for the 13 nm thin film, selected as representative example.

## Concluding remarks

In this paper we have reported about the first synthesis by means of rf-sputtering of optimally doped $La_{0.85}Na_{0.15}MnO_3$ thin films. We selected three different substrates and for each of them three different thickness in order to look at their effect on the physical properties of the materials, particularly on the most important for applicative purposes, *i.e.* the magnetoresistance.

First of all, the synthetic procedure here applied showed to be suitable to produce well oriented thin films of sodium-doped lanthanum manganites. The structural data confirmed that films grown onto STO (100) are under tensile strain while the ones grown onto NGO (110) are under a compressive strain with an in-plane mismatch of the same order of magnitude but of opposite sign. Finally for the series on NGO (100) an orthorhombic structure is stabilized by the oriented growth of the thin film. In all cases, the thickness increase leads to a partial removal of the strain induced by the substrate as can be confirmed (in particular for the thin films grown on STO (100) and NGO (110)) by the progressive shift of the lattice constants towards the bulk values.

The transition temperatures in the *R vs. T* curves for all the films considered are lower with respect to the bulk material with the same nominal composition[18] for which $T_{S-M}$ is around 300 K (as clearly stressed above, the actual average Mn valence is the same in all the films and in the bulk materials). This behavior is usually observed for the manganite thin films and can have many different sources[1]. Moreover, a size effect of the film grains should be taken into account particularly for what concerns the lowering of transition temperatures with respect to the bulk (micro-sized) material. The behavior of the transport properties for the STO (100) series is interesting since it reflects the role of the various *degrees of freedom* actually present when dealing with thin films with respect to the bulk materials. The thinnest film (13 nm) has a $T_{S-M}$ of 248 K, the maximum $T_{S-M}$ value of this series (260 K) pertains to the 25 nm film and it reduces by increasing thickness. The lower $T_{S-M}$ value of the 13 nm film with respect to the 25 nm film can be directly connected to a proximity effect on the transport properties of an insulating dead layer which is usually found in these thin films. For epitaxial thin films of Ca-doped LaMnO$_3$ grown on STO (100) this layer was found to be around 7 nm[38]. In the dead layer the magnetic interaction is non-FM and induces the formation of regions with localized charge carriers. Of course this dead layer, whose exact nature and origin is still source of debate, is close to the next metallic and ferromagnetic layer. We can expect then that near the boundaries also some distortion of the MnO$_6$ octahedra is present giving origin to a partial values distribution of the Mn-O distances and Mn-O-Mn angles in the FM region[38]. This should cause a reduction of both $T_C$ and $T_{S-M}$ but the effect on the last should be also more evident due to the possible formation of highly localized states near the top and the bottom of the conduction band and more generally to the introduction of trapping centers by structural disorder. Direct evidence of a greater effect of this electronic and structural disorder on the transport properties can be argued by looking at the $T_C$ trend which, opposite to the $T_{S-M}$ trend, follows a monotonic behavior as outlined in the Results and Discussion section. Something analogous was also already pointed out by our previous works on bulk materials[9,18].

Further increasing the film thickness in the STO series leads to the appearance of two distinct transitions in the resistance curves: one located at the same temperature as the 25 nm film (260 K) and another one to a lower temperature (182 K). This result can be accounted for the strain relaxation effect which can cause the formation of distinct layers with different strains and thus different transition temperatures. Also in this case the strain effect is more clearly observed in the transport properties since just one magnetic transition is detected at a temperature *incidentally* corresponding to the mean value of the two $T_{S-M}$ transitions.

The experimental results for the film deposited on NGO (110) bring in prominence the presence of a dead layer also for the LNMO phase since the thinnest film of this series (11 nm) is completely insulating. We remark that this feature has been previously observed only for LCMO and LSMO[17,38-41] materials while this is the first report about an insulating dead layer for an alkali-doped manganites thus confirming this aspect as a general rule for manganites thin films.

For the intermediate thickness of thin films on NGO (110) we achieved the highest transition temperature (285 K) coupled to an MR response of 70%. As already told in the Introduction, *MR* values of this order have been previously obtained in manganites thin films only for lower temperature in the LCMO phase, while higher $T_{S-M}$ have been reported for the LSMO phase but coupled to very low *MR* values.

Also for the films on NGO (110) an increase of the thickness reduces $T_{S-M}$, mainly as a consequence of relaxation effects even though the *MR* response is now pushed to more than 90%. Anyway, the concomitant increase of the *MR* effect by reducing transition temperature has been connected to a higher sensitivity of the systems to magnetic field variation as the transition temperature decreases due to the lower difference in the resistance between a metallic phase and a semiconductor-like phase[42].

The last series of films grown, *i.e.* those on NGO (100), showed to be completely insulating with extremely high resistance values. Only for the thickest film (82 nm) a small effect of the 7 T field was observed (see Figure 5d). In this case the oriented growth of the manganite films imposes an orthorhombic deformation of the structure which inhibits the ferromagnetic coupling and the charge carrier transport. When the thickness increases, the usual tendency towards strain relaxation and possible loss of orientation due to extended defects can allow the formation of metal-like FM clusters which induce a general reduction of the resistance and a small *MR* effect. The phase stabilized by the NGO (100) substrate could be a charge-ordered one as already observed in analogous systems[43] even though for very thin films. Unfortunately, the paramagnetic nature of the substrate prevents an analysis of the magnetic properties of the film grown on the NGO substrate.

To conclude, in this paper we have reported the first study of rf-sputtered sodium-doped lanthanum manganites thin films. We have shown the possibility to tune the physical properties in the deposited thin films by changing the kind of substrate, its orientation and the film thickness. Even though this is something already known from the previous literature few works reported data about structural, magnetic and electrical properties for different substrates and thickness for a selected compound; moreover, this represent the first thorough work on an alkali-doped manganite.

Among the several possible parameters that can influence these properties of manganites films it has argued that the film strain plays a crucial role. However, additional parameters (as relaxation-induced disorder, unhomogeneity, morphology features) must be taken into account in order to depict a more clear *scenario* for these complex oxides. The results of our work have shown a way to optimize the $T_{S-M}$ and the *MR* response: by looking at a more suitable starting material with respect to the most common ones and by tuning the physical properties with a convenient choice of substrate type, orientation and film thickness an *MR* response of 70% coupled to a transition temperature of 285 K has been found.

**Acknowledgments**

Financial support from the Italian Ministry of Scientific Research (MIUR) by PRIN Projects (2002) is gratefully acknowledged. Dr. Oscar Barlascini is acknowledged for having performed the EPR measurements. Dr. Elza Bontempi is acknowledged for experimental support (University of Brescia).

**Figure captions**

**Fig. 1.** XRD $\theta$-$2\theta$ scans for the 13 nm LNMO thin film on STO (100) (a); the 82 nm LNMO thin film on NGO (110) (b) and the 82 nm LNMO thin film on NGO (100) (c).

**Fig. 2.** AFM images for the as-deposited 33 nm LNMO film on NGO (110). See details in the text.

**Fig. 3.** AFM images for the annealed 33 nm LNMO film on NGO (110). See details in the text.

**Fig. 4.** *R* vs. *T* curves at *H* = 0 (solid line) and at *H* = 7 T (dash-dot line) and *MR* curves (empty circles) for the annealed LNMO thin films deposited on STO (100) of thickness 13 nm (a), 25 nm (b) and 70 nm (c).

**Fig. 5.** *R* vs. *T* curves at *H* = 0 (solid line), *H* = 1 T (dashed line) and *H* = 7 T (dash-dot line) and *MR* curves (empty circles) for the annealed LNMO thin films deposited on NGO (110) of thickness 11 nm (a), 33 nm (b) and 82 nm (c). Figure 5d refers to LNMO thin films deposited on NGO (100) where the triangles represent the curves at 0 T while the diamonds refer to the curves at 7 T. Thickness for this series are directly reported in the Figure.

**Fig. 6.** *M* vs. *T* curves at 1000 Oe (magnetic field parallel to sample surface) for LNMO thin films deposited on STO (100) of thickness 13 nm (circles), 25 nm (squares) and 70 nm (triangles).

**Fig. 7.** Hysteresis loops at 2K for LNMO thin films deposited on STO (100) of thickness 13 nm (circles), 25 nm (squares) and 70 nm (triangles). The inset shows an enlarged image of the same curves.

**Fig. 8.** Temperature dependence of the EPR signal of the 13 nm LNMO thin film deposited on STO (100) obtained by applying the magnetic field parallel to the sample surface. Signals at 278 K, 294 K and 328 K have been amplified 2, 5 and 10 times, respectively.

**Fig. 9**. Temperature dependence of the EPR resonant field of the 13 nm LNMO thin film deposited on STO (100) for magnetic field applied both parallel (open symbols) and perpendicular (full symbols) to the sample surface.

**Table caption**

Substrate type, film thickness (nm), out-of-plane parameter (Å), transition temperatures in the *R vs. T* curves, activation energies (meV), maximum of MR response (%) and temperature of the MR maxima.

**Table 1**

| *Substrate* | *Thickness* (nm) | *Out-of-plane parameter* (Å) | $T_{\text{S-M}}$ (K) | $E_a$ (meV) | *MR (%)* 7 T | *MR peak* 7 T (K) |
|---|---|---|---|---|---|---|
| STO (100) | 13 | 3.851 | 248 | 94.2 | 77.9 | 240 |
|  | 25 | 3.878 | 260 | 95.7 | 78.2 | 252 |
|  | 70 | 3.876 | 260, 182 | 129 | 45; 37 | 259; 178 |
| NGO (110) | 11 | - | - | 155 | - | - |
|  | 33 | 3.898 | 285 | 71.6 | 70 | 271 |
|  | 82 | 3.892 | 256 | 114 | 91 | 240 |
| NGO (100) | 11 | 3.877 | - | 221.1 | - | - |
|  | 33 | 3.879 | - | 188 | - | - |
|  | 82 | 3.884 | - | 154.2 | - | - |

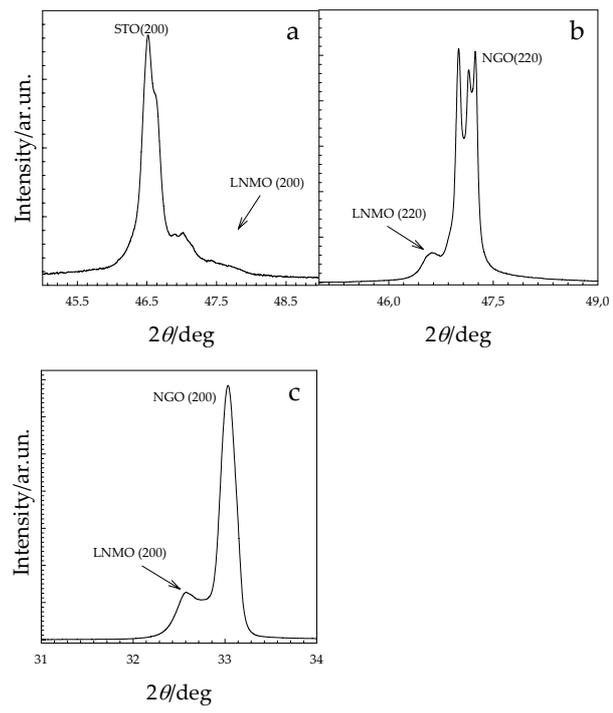

Figure 1

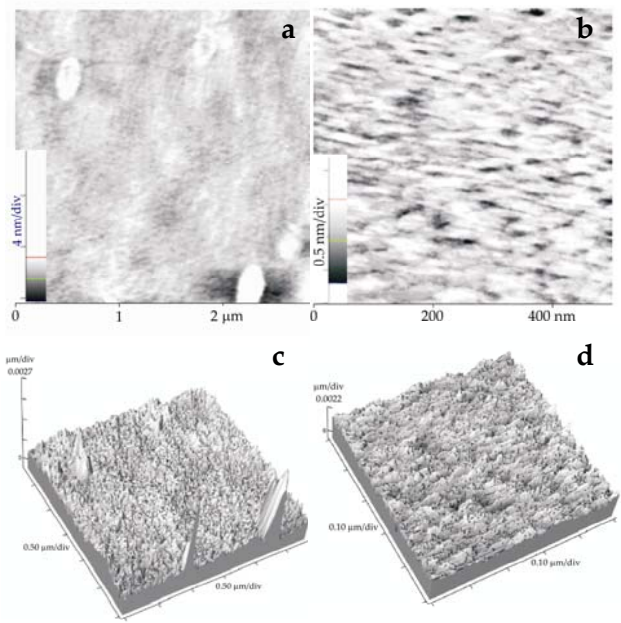

Figure 2

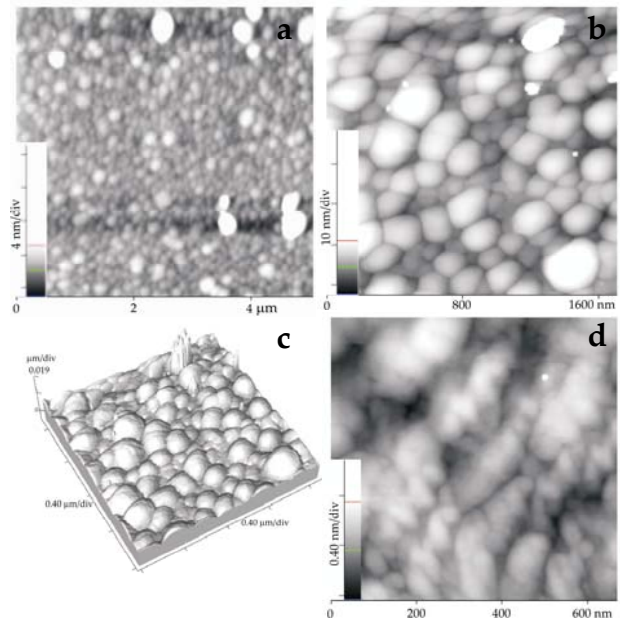

Figure 3

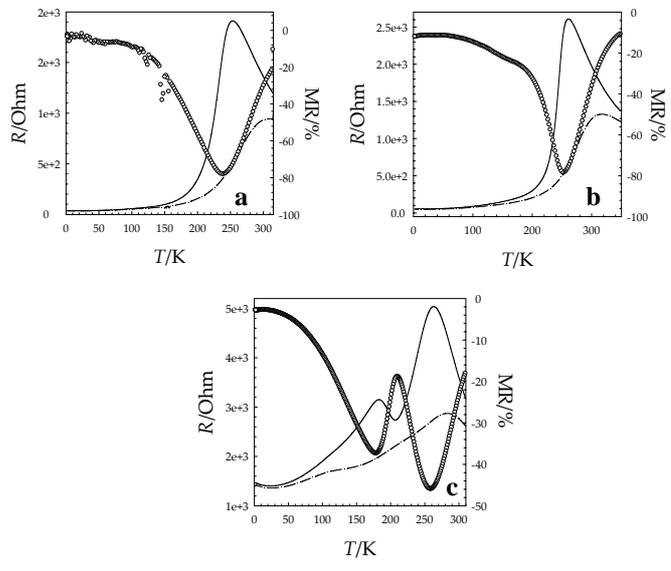

Figure 4

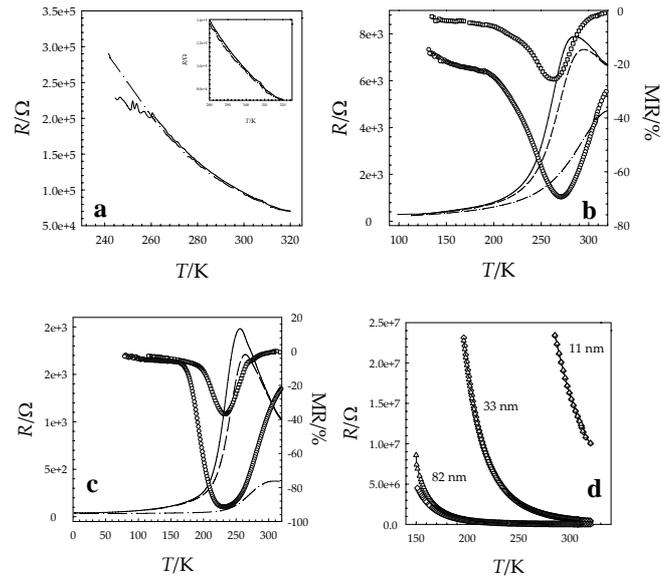

Figure 5

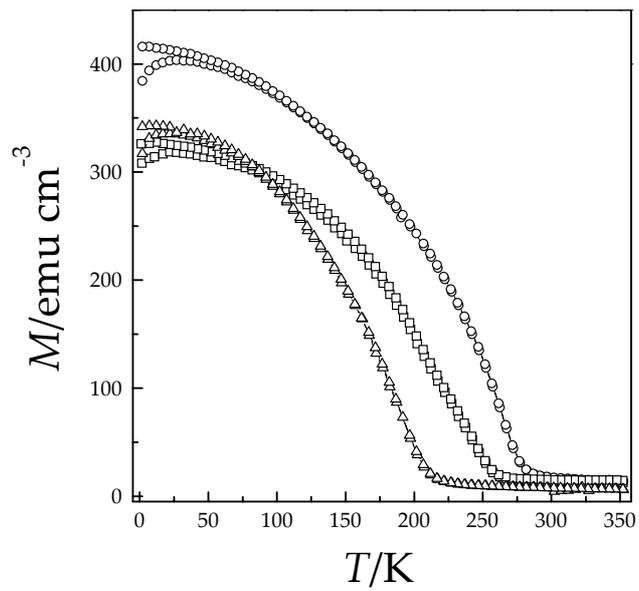

Figure 6

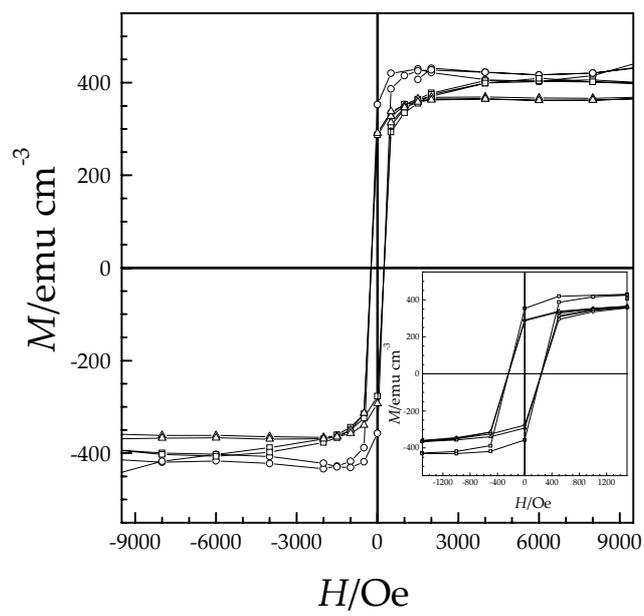

Figure 7

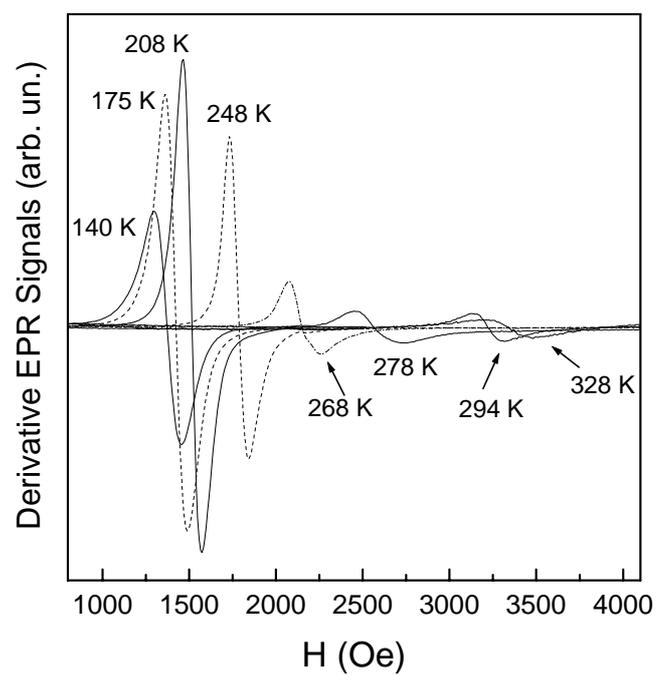

Figure 8

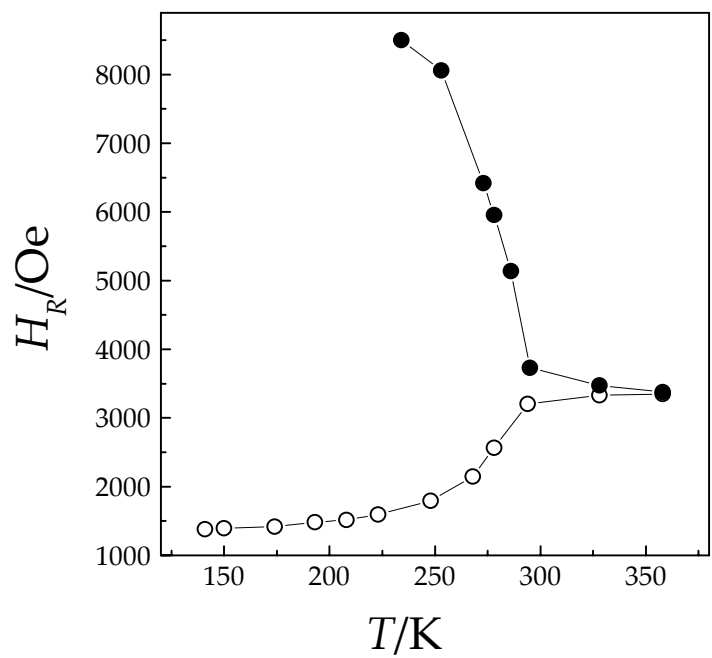

Figure 9